\documentclass[conference]{IEEEtran}
\IEEEoverridecommandlockouts

\usepackage{fixmath}
\usepackage{gensymb}
\usepackage[center]{caption}
\usepackage{subcaption}
\usepackage{multirow}
\usepackage[english]{babel}
\usepackage{cite}
\usepackage{epstopdf}
\usepackage{amsmath,amssymb,amsthm,mathrsfs,amsfonts,dsfont}
\usepackage{algorithmic}
\usepackage{graphicx}
\usepackage{graphics}
\usepackage{color}
\usepackage{textcomp}
\usepackage{balance}
\usepackage[top=0.75in,bottom=1in,left=0.675in,right=0.675in]{geometry}

\newcommand{\site}{\boldsymbol{s}}
\def\BibTeX{{\rm B\kern-.05em{\sc i\kern-.025em b}\kern-.08em
		T\kern-.1667em\lower.7ex\hbox{E}\kern-.125emX}}
\begin{document}
	
	\title{3D Beamforming based Dynamic TDD Interference Mitigation Scheme}

	\author{\IEEEauthorblockN{Jalal Rachad$~^{*+}$, Ridha Nasri$~^*$, Laurent Decreusefond$~^+$}
		\IEEEauthorblockA{$~^*$Orange Labs: Direction of Green transformation, Data knowledge, traffic and resources Modelling,\\ 40-48 avenue de la Republique
			92320 Chatillon, France\\ 
			$~^+$LTCI, Telecom Paris, Institut Polytechnique de Paris,\\
			 75000, Paris, France\\
			Email:$^{*}$jalal.rachad@ieee.org, $~^*$ridha.nasri@orange.com,
      $^+$Laurent.Decreusefond@mines-telecom.fr}
}		
	
	\maketitle
		\begin{abstract}
			
		Dynamic Time Division Duplexing (D-TDD) allows cells to accommodate asymmetric traffic variations with high resource assignment flexibility. However, this feature is limited by two additional types of interference between cells in opposite transmission direction: downlink (DL) to uplink (UL) and UL to DL interference. Therefore, using this mode with macro-cell deployments requires interference mitigation techniques to reduce the strong DL to UL interference. 3D beamforming is an efficient technique that minimizes interference and enhances performance by exploiting a large 2D array of antennas intelligently. Combining D-TDD and 3D beamforming can make D-TDD feasible for macro-cells. The aim of this work is to provide a 3D beamforming analytical model in a D-TDD based macro-cells' deployment where beamforming horizontal and vertical radiation patterns depend on the spatial distribution of random users' locations. We evaluate interference in terms of Interference to Signal Ratio (ISR). We show that the cumulative ISR can be written in terms of convergent series and its expectation is an almost sure convergent series. Different numerical results are presented to justify the applicability of this scheme.   	
			
	    \end{abstract}

	
\section{Introduction}

%As the number of mobile users and the volume of data traffic are expected to continue increasing in the upcoming years, future mobile cellular networks need to support this proliferation through upgrading their features and key technologies. 

Time division duplex (TDD) is expected to be a key feature for the upcoming fifth generation (5G) of cellular networks. It offers more advantages than frequency division duplex (FDD) mode in terms of capacity enhancement, flexibility and implementation adequacy with other features such as full dimension multiple input multiple output (FD-MIMO). A variant operational mode of TDD is D-TDD that provides more flexibility in resource assignment. D-TDD has been proposed as a solution to deal with DL and UL traffic asymmetry since it is based on instantaneous traffic estimation. However, this duplexing mode is severely limited by a strong mutual interference between the UL and DL transmissions, called cross slot interference (CSI), because those two directions share the same frequency band. There is two types of CSI: DL to UL (impact of the DL other cells interference on the UL signal received by the studied cell) and UL to DL (impact of the UL mobile users transmission, located in other cells, on the DL signal received by a mobile user located in the studied cell). Most interference mitigation schemes, such as cell clustering and Further enhanced Inter Cell Interference Coordination (FeICIC) \cite{rachad2018interference}, are dedicated to minimize D-TDD interference in heterogeneous networks. Recently, the applicability of massive multiple antenna technologies, such as FD-MIMO and 3D beamforming, with D-TDD has drawn the attention of telecommunication actors. Actually, 3D beamforming consists in creating a signal beam between the transmitter and the receiver, in the horizontal and the vertical dimensions, by using a two-dimensional array of active antennas. It enhances the signal strength at the receiver and minimizes interference level so that high average data rate and high spectral efficiency can be achieved. Applying 3D beamforming can effectively reduce the strong DL to DL and DL to UL interference impact and thus make D-TDD feasible for macro-cell deployments.\\

D-TDD was the subject matter of many works in literature. In \cite{rachad2018interference}, it has been proposed  an analytical model for interference tractability in D-TDD system. The explicit formulas of $ISR$, covering the four D-TDD interference scenarios, have been derived. Authors showed that D-TDD can be used in favor of the DL transmission direction. However, the UL transmission is completely limited by DL to UL interference. Similarly in \cite{khoryaev2012performance}, through system level simulations of a D-TDD based small-cells network, authors have reached the same conclusions as \cite{rachad2018interference}. An interesting D-TDD interference tractability approach in a particular small-cells' architecture known as phantom cells, based on stochastic geometry, can be found in \cite{yu2015dynamic}. Likewise, in \cite{kulkarni2017performance} authors have provided a comparison between static and dynamic TDD in millimeter wave (mm-wave) cellular network, in terms of $SINR$ distributions and mean rates, considering synchronized and unsynchronized access-backhaul. On the other hand, FD-MIMO has been introduced as an efficient feature to enhance mobile network performance in terms of users' throughputs and spectral efficiency. It consists in arranging a large number of antennas in a 2D array which enables to use 3D beamforming \cite{nadeem2018elevation}. In our recent work \cite{rachad20193d}, we have proposed a 3D beamforming scheme where antenna horizontal and vertical radiation patterns depend on the spatial distribution of users' locations. System level simulations have shown that this feature reduces significantly interference and enhances the $SINR$ and thus users throughput in DL. Furthermore, the marriage between FD-MIMO and D-TDD can make D-TDD feasible for macro-cell deployments. For instance in \cite{huang2018technical}, based on random matrix theory, authors have shown that equipping BSs with a large number of antennas removes effectively the DL to UL interference in macro-cell deployments.\\

The main contribution of this paper is to provide an analytical 3D beamforming model that can be applied to D-TDD based systems. We focus on the explicit analysis of $ISR$ by covering different interference scenarios in DL and UL transmission directions. We show that the average DL and UL $ISRs$ can be expressed as an almost sure convergent series of independent random variables. Finally we analyze, through system level simulations, the applicability of the proposed 3D beamforming scheme in the context of D-TDD and also, we compare performance to that of Static TDD (S-TDD) based system. To the best of our knowledge, this is the first work providing a practical analytical model that combines D-TDD with 3D beamforming.\\

The remainder of this paper is organized as follows: in section II, we describe the network model, D-TDD model and the 3D beamforming scheme. Section III is devoted to provide some analytical results regarding interference characterization. Simulations results are given in section IV. Section V concludes the paper.

\section{System model and notations}

\subsection{Network model}

We consider a regular tri-sectorized hexagonal network denoted by $\Lambda$ with an infinite number of sites $\site$  having an inter-site distance denoted by $\delta$. For each site $\site$ $\in$ $\Lambda$, there exists a unique (m,n) $\in$ $\mathbb{Z}^2$ such that $\site = \delta( m + ne^{i\frac{\pi}{3}})$. We denote by $\site_0$ the serving cell located at the origin of $\mathbb{R}^2$. Unlike regular hexagonal network with omni-directional antennas, BSs of sites are located at the corner of the hexagons. All BSs have the same height $l_b$,  transmit with the same power level $P$ and assumed to have directional antennas covering, each one, a hexagonal sector identified by $c \in \{1,2,3 \}$. The azimuths of antennas $\vartheta_c$ in which the radiation is at its maximum are taken relative to the real axis such that $\vartheta_c=\frac{\pi}{3}(2c-1)$. So the azimuth of the first sector of each site has an angle of $\frac{\pi}{3}$ with the real axis relatively to the location of $\site$.\\

We consider a typical mobile served by the first sector ($c=1$) of $\site_{0}$. Its location is denoted by $z_0$ such that $z_0 = re^{i\theta}$ where ($r$, $\theta$) are the polar coordinates in the complex plane. We denote also by $z_{\site,c}$ the location of a mobile served by a sector $c$ of a site $\site$ $\in$ $\Lambda^{*}$, where $\Lambda^{*}$ is the lattice $\Lambda$ without the typical site $\site_0$. Locations $z_{\site,c}$ are written in the complex plane by $z_{\site,c} = \site + r_{\site,c} e^{i\theta_{\site,c}}$, where $r_{\site,c}$ and $\theta_{\site,c}$ represent respectively the distance and the angle (complex argument) between $z_{\site,c}$ and $\site$.

\subsection{Dynamic TDD model}

\begin{figure}[tb]
	\centering
	\begin{subfigure}[b]{0.35\textwidth}
		\includegraphics[width=1\linewidth]{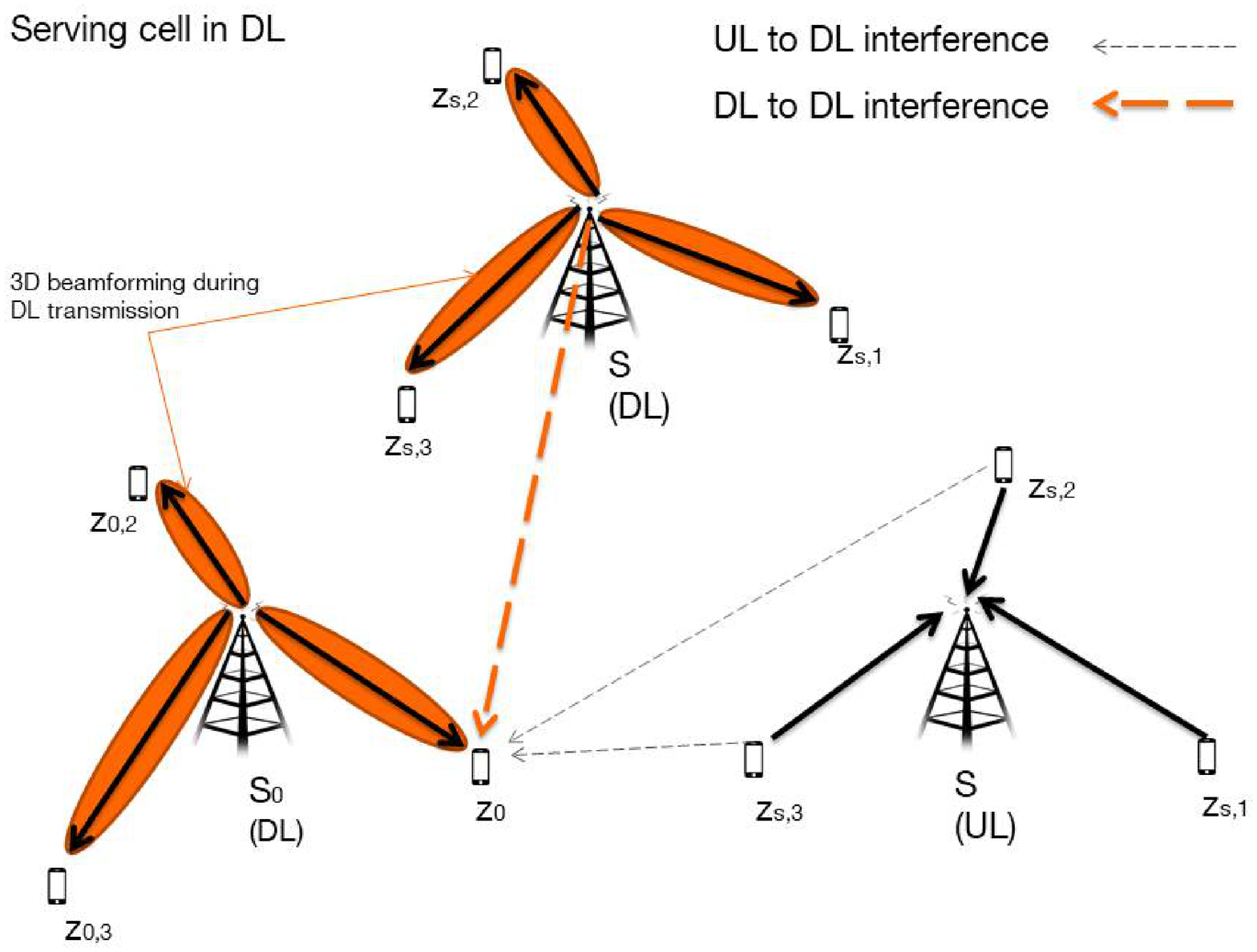}
		\caption{}
		\label{fig:Ng1} 
	\end{subfigure}
	
	\begin{subfigure}[b]{0.35\textwidth}
		\includegraphics[width=1\linewidth]{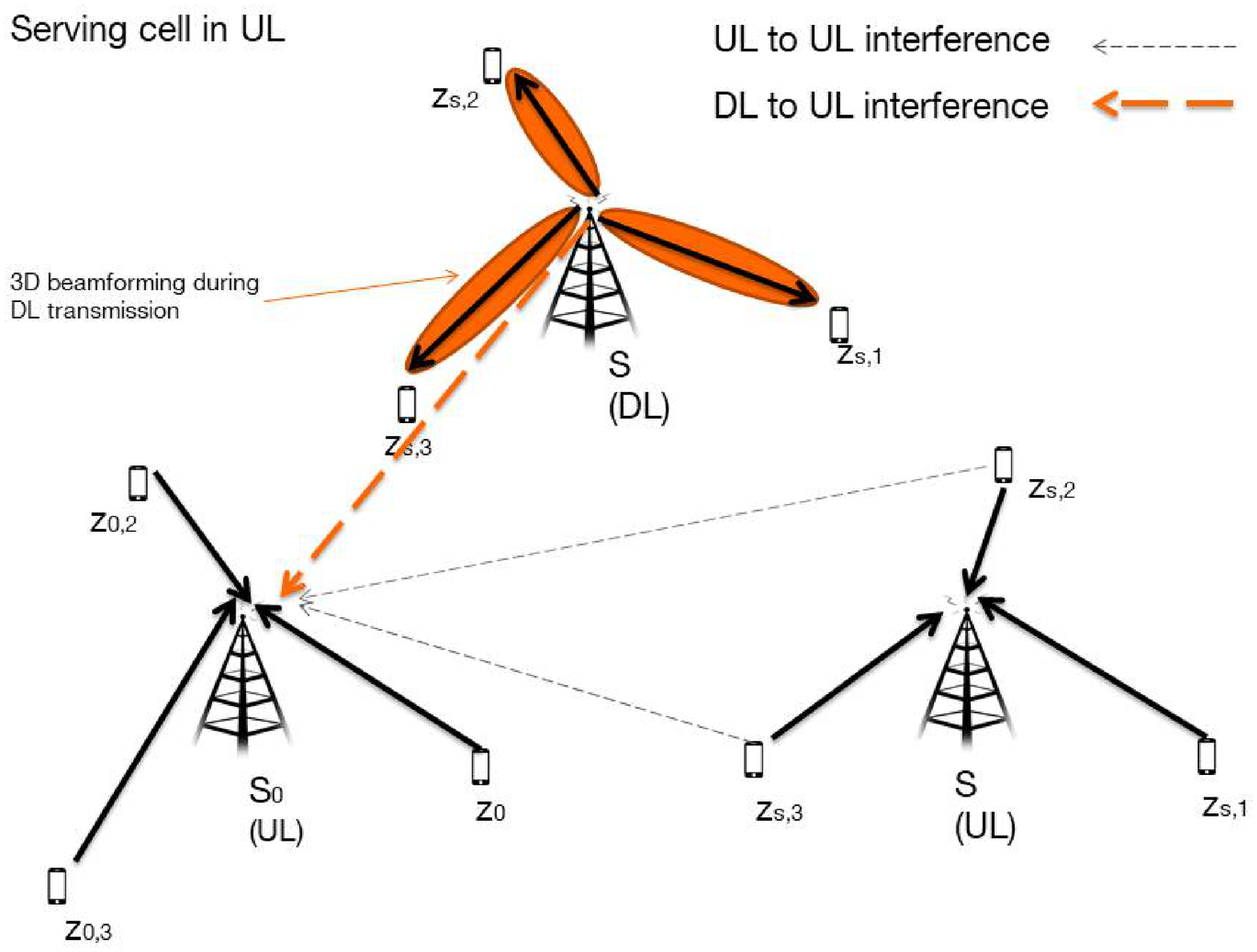}
		\caption{}
		\label{fig:Ng2}
	\end{subfigure}
	
	\caption{D-TDD interference scenarios: (a) the serving cell is operating in DL; (b) the serving cell is operating in UL}
	\label{D-TDD illustration}
\end{figure}

To model the D-TDD system, we assume that all cells initially operate synchronously in DL or UL. This setup can be considered as a baseline scenario characterizing performance of existing S-TDD systems. After a period of time, it is assumed that all cells randomly select UL or DL transmission directions based on traffic estimation. Also, we assume that the transmission direction is the same in the three sectors of each site. Four types of interference henceforth appear depending on the transmission cycle of each site: \textit{i}) when the serving cell transmits to a given mobile location, DL useful signal is impacted by interference from DL BSs and UL mobiles' signals; \textit{ii}) when the serving cell operates in UL, the received UL signal is interfered by DL signals from BSs and UL signals from mobiles (Fig. \ref{D-TDD illustration}). It is considered hereafter that the scheduler does not allocate the same spectral resources to different mobile users in one cell at the same time (e.g., TD-LTE scheduling). So, intra-cell interference is not considered. Therefore, in a given cell,  we consider that during a sub-frame of interest (i.e., when D-TDD is activated), there is one active transmission whether in DL or UL with full-buffer traffic model. Additionally, to characterize the transmission direction of each site $\site$, we consider two Bernoulli RVs $\beta_d(\site)$ and $\beta_u(\site)$ such that $\mathbb{P}(\beta_d(\site)=1)=\alpha_d$ and $\mathbb{P}(\beta_u(\site)=1)=\alpha_u$. $\beta_d(\site)$ ($\beta_u(\site)$) refers to the DL (UL) transmission cycle of a site $\site$ during a D-TDD sub-frame. It is important to mention that a site $\site$ cannot be in DL and UL during the same TTI. Hence,  we assume that  $\beta_d(\site)=1-\beta_u(\site)$. This means that $\alpha_{d}=1-\alpha_{u}$.

%Furthermore, it has been shown in \cite{rachad2018interference} that D-TDD performs better during the DL transmission cycle of a cell than the S-TDD. However, the UL transmission is severely impacted by DL to UL interference. 3D beamforming appears to be an efficient solution to reduce the strong interference coming from BSs' signals. Nevertheless, applying this feature requires beam and frequency resources management mechanisms between interfering BSs. We assume in the sequel that, during the sub-frames where cells are allowed to use dynamic resources allocation, transmitting cells in DL adopt 3D beamforming to serve DL users.

\subsection{3D beamforming scheme}

 BSs are equipped with directional antennas with sectorized gain pattern. To model the 3D beamforming,  we assume that each antenna has a directional radiation that can be described by two planar patterns: the horizontal and the vertical one denoted respectively by $H$ and $V$. We define the antenna radiation for each pattern in the linear scale by a $2\pi$-periodic function, according to Mogensen model \cite{mogensen1997preliminary} and \cite{rachad20193d}, as follows

 \begin{align}
 &H(\alpha)= [\cos(\alpha)]^{-2w_h} \\
 &V(\phi)=[\cos(\phi)]^{-2w_v},
 \label{Mogensen}
 \end{align} 
 with $w_h=\frac{ln(2)}{\ln(\cos(\frac{\theta_{h3dB}}{2})^2)}$ and $w_v=\frac{ln(2)}{\ln(\cos(\frac{\theta_{v3dB}}{2})^2)}$. $\theta_{h3dB}$ and $\theta_{v3dB}$ are respectively the horizontal and the vertical half power beam widths.
 
 The beamforming antenna radiation pattern received in a mobile location $z_0$ from an interfering site $\site$ is defined by
 
 \begin{equation}
 G_{\site}(z_0)=\sum_{c=1}^{3} H(\alpha_{\site,c})V(\phi_{\site,c}),
 \label{gain}
 \end{equation} 
 with $\alpha_{\site,c}$ is the angle between the mobile $z_0$ orientation and the beam axis directed to a mobile $z_{\site,c}$ in the horizontal plane and $\phi_{\site,c}$ is the angle between the beam direction in the vertical plane and the mobile $z_0$. The angle $\alpha_{\site,c}$ can be expressed, based on the complex geometry, as
 
 \begin{equation}
 \alpha_{\site,c}=  \psi(z_0,\site) - \theta_{\site,c}, 
 \label{alpha}
 \end{equation} 
 where $ \psi(z_0,\site)=arg(z_0-\site)$ and $\theta_{\site,c}=\psi(z_{\site,c},\site)$ is the complex argument of $z_{\site,c}$ relatively to $\site$. Each $\theta_{\site,c}$ is assumed to be uniformly distributed in the interval $[\vartheta_c-\frac{\pi}{3},\vartheta_c+\frac{\pi}{3}]$, thus using a linear transformation of the RV $\theta_{\site,c}$, we can easily prove that the angle $\alpha_{\site,c}$ is a RV uniformly distributed in the interval $[ \psi(m,\site)-\frac{2\pi}{3}c, \psi(m,\site)+\frac{2\pi}{3}(1-c)]$.\\
 
 Similarly for the vertical dimension,  the angle $\phi_{\site,c}$ can be expressed as
 \begin{equation}
 \phi_{\site,c}= atan(\frac{l_b}{|z_0-\site|})- \tilde{\phi}_{\site,c},
 \label{phi1}
 \end{equation}
 with $\tilde{\phi}_{\site,c}=atan(\frac{l_b}{r_{\site,c}})$  refers to the antenna downtilt, which is variable in this case.\\
 
 The distance $r_{\site,c}=|z_{\site,c}-\site|$ between a mobile $z_{\site,c}$ and $\site$ varies between $0$, when $z_{\site,c}$ is close to $\site$ location, and $\frac{2\delta}{3}$ when $z_{\site,c}$ is located at the far edge of a hexagonal sector. It can be characterized by using the expression of the horizontal antenna radiation pattern that covers a whole hexagonal sector of $\site$. Thus, $r_{\site,c}$ is described by a RV varying between $0$ and $\frac{2\delta}{3}U(\theta_{\site,c}-\vartheta_c)$, with
 
 \begin{equation}
 U(\theta_{\site,c}-\vartheta_c)=[\cos(\theta_{\site,c}-\vartheta_c)]^{-2w_h}
 \end{equation}
 is the horizontal antenna radiation pattern that covers a hexagonal sector (i.e., the half power beam width $\theta_{h3dB}=65\degree$).
 
 Based on that, a mobile will be located at the far edge of a sector when the angle between $z_{\site,c}$ and $\site$ is equal to the antenna azimuth in which the radiation is at its maximum. Also, we assume in the remainder that $r_{\site,c}$ is a RV uniformly distributed on the interval $[0,\frac{2\delta}{3}U(\theta_{\site,c}-\vartheta_c)]$.
 
\subsection{Propagation model}

To model the wireless channel, we consider the standard power-law path loss model based on the distance between a mobile $z$ and a BS $\site$ such that the path loss $L(\site,z)$ is given by

\begin{equation}
L(\site,z)=a|\site-z|^{2b},
\label{pathloss}
\end{equation}
with $2b$ is the path loss exponent and $a$ is a propagation factor that depends on the type of the environment (Indoor, Outdoor...).

In addition to the path loss, the received power by a mobile depends on the random channel effects, especially shadowing and fast fading. Shadowing refers to the attenuation of the received signal power caused by obstacles obstructing the propagation between the transmitter and receiver. We model the shadowing effect between a transmitting node $t$ and a receiving one $r$ by a log-normal RV $\chi(t,r)= 10^{\frac{Y(t,r)}{10}}$, where $Y(t,r)$ is a normal RV with mean $\mathbb{E}(Y(t,r))=0$ and variance $\sigma^2$. This sequence of RVs are assumed to be independent and identically distributed for all $(t,r)$.

On the other hand, fast fading random model is not considered in this paper. Its effect can be compensated through link level performing that maps the $SINR$ to the throughput ($Th$). Also, for an AWGN (Additive Gaussian Noise Channel), Shannon's formula provides the relation between $SINR$ and $Th$. Hence, the fast fading effect can be compensated by using a modified Shannon's formula to have $Th=K_1 log_2(1+K_2 SINR)$, with $K_1$ and $K_2$ are constants calibrated from practical systems \cite{mogensen2007lte}.

Additionally, for the UL transmission direction, power control is applied to PUSCH (Physical Uplink Shared Channel) channel in order to set the required mobile transmitted power. In this paper, it is modeled by the fractional power control (FPC) model, i.e., the path loss is partially compensated by the power control \cite{castellanos2008performance}. The transmitted power by the mobile location $z$ to its serving cell $\site$ is then written as
\begin{equation}
P_t(\site,z)= P^*(\site) \left(\left|z - \site \right|^{2b}\right)^{k}
\label{power control}
\end{equation}
where $P^*$($\site$) is the target cell specific power and $k$ $\in$ [0 1] is the power control compensation factor. When $k=1$ the power control scheme totally indemnifies the path loss in order to reach the target power $P^*(\site)$. For the case $0<k<1$ the path loss is partially compensated and mobile users in cell edge create less interference because their transmitted power is reduced. Without loss of generality, we consider that $P^*(\site)$ is the same for all the cells and we denote it by $P^*$. Also we assume that $P$ and $P^*$ include the BSs and mobiles antenna gain.\\

Therefore, the received power from a BS $\site$, transmitting with a power level $P$, at a mobile location $z_0$ is expressed by

\begin{equation}
P_r(z_0,\site)=\frac{P  G_{\site}(z_0) \chi(\site,z_0)}{L(\site,z_0)},
\label{receivedpowerbs}
\end{equation}   
with $G_{\site}(z_0)$ is the antenna beamforming radiation pattern from $\site$ received at the mobile location $z_0$.

Similarly, the received power from an interfering mobile $z_{\site,c}$, located at the sector $c$ of a site $\site$, received at a mobile location $z_0$ is

 \begin{equation}
 P_r(z_0,z_{\site,c})=\frac{P_t(\site,z_{\site,c}) \chi(z_{\site,c},z_0)}{L(z_{\site,c},z_0)},
 \label{receivedpowermobile}
 \end{equation}  

\section{Interference characterization}
 
 We define the Interference to Signal Ratio $ISR$ as the received power from an interfering source (interfering mobile or BS) divided by the useful signal power received by $z_0$ (or $\site_0$) from the serving cell (or UL transmitting mobile $z_0$). In DL transmission, the mobile $z_0$ receives interference from other BSs operating in DL and from mobiles transmitting in UL. Hence, we define the DL $ISR$ denoted by $\mathcal{I}_{DL}(z_0)$ by
 
 \begin{equation}
 \mathcal{I}_{DL}(z_0)= D_{\downarrow}(z_0) + D_{\uparrow}(z_0)
 \label{ISRDLdef}
 \end{equation}
 where $D_{\downarrow}(z_0)$ is the DL to DL $ISR$ and $D_{\uparrow}(z_0)$ is the UL to DL $ISR$. $D_{\downarrow}(z_0)$ and $D_{\uparrow}(z_0)$ can be defined respectively as follows

 	\begin{align}
 	&D_{\downarrow}(z_0)= -1 + \sum_{\site \in \Lambda} \beta_d (\site) G_{\site}(z_0) |\site - z_0|^{-2b} r^{2b} \tilde{\chi}(\site,z_0)    \\
 	&D_{\uparrow}(z_0)= \sum_{\site \in \Lambda^*}  \beta_u (\site) \sum_{c=1}^{3} \frac{ r^{2b} P^* r_{\site,c}^{2bk} \tilde{\chi}(z_{\site,c},z_0)}{P |z_{\site,c} -z_0 |^{2b}},  
 	\end{align}   
  where $\tilde{\chi}(t,z_0)=10^{\frac{\tilde{Y}(t,z_0)}{10}}$ is a log-normal RV representing the ratio of $\chi(t,z_0)$ ( $t= \site$ or $t=z_{\site,c}$) and $\chi(\site_0,z_0)$ (The ratio of two log-normal RVs is a log-normal RV). $\tilde{Y}(t,z_0)$ is a Normal RV with mean 0 and variance $\tilde{\sigma}^2$.\\
  
  $\mathcal{I}_{DL}(z_0)$ is an infinite sum of independent positive RVs not identically distributed. As far is known, the distribution of a sum of log-normal RVs is not exactly determined but there exists some approximations that can be found in the probability theory literature. Fenton-Wilkinson's  approach \cite{fenton1960sum} appears to be a convenient approximation to deal with this sum of log-normal RVs. It consists in approximating a sum of log-normal RVs by a new log-normal RV by matching the mean and the variance. Hence conditionally on the mobile location $z_0$ and according to Theorem 9.2.a of \cite{jacod2012probability}, (12) becomes
  
  \begin{align}
  &\mathbb{E}[ D_{\downarrow}(z_0)]= -1 + \alpha_d e^{\frac{\ln^2(10)\tilde{\sigma}^2}{200}}\sum_{\site \in \Lambda} \mathbb{E}[G_{\site}(z_0)] |\site - z_0|^{-2b} r^{2b}.
  \end{align}
  
   In \cite{nasri2016analytical}, it has been shown that $\sum_{\site \in \Lambda^*} r^{2b}|\site-m|^{-2b}$ is a convergent series on $x=\frac{r}{\delta}$ that can be approximated as follows 
  
  \begin{equation}
  \sum_{\site \in \Lambda^*} r^{2b}|\site-z_0|^{-2b}\approx \frac {6x^{2b}}{\Gamma(b)^{2}}\sum_{h=0}^{+\infty}\frac{\Gamma(b+h)^{2}}{\Gamma(h+1)^{2}}\omega(b+h)x^{2h} 
  \label{ridha}
  \end{equation}
  \vspace*{0.3cm}
  \begin{equation}
  \text{ where }\omega(z) = 3^{-z}\zeta(z)\left( \zeta(z,\frac{1}{3})-\zeta(z,\frac{2}{3})\right), \nonumber
  \label{omega}
  \end{equation}
  with $\zeta(.)$ and $\zeta(.,.)$ are respectively the Riemann Zeta and Hurwitz Riemann Zeta functions \cite{abramowitz1964handbook}.\\
  
  Since $G_{\site}(z_0)\leq 1$, we have $\mathbb{E}(G_{\site}(z_0))< 1 $. By using (\ref{ridha}), we can easily prove that $\mathbb{E}[ \mathcal{I}(z_0)] < \infty$. Hence, according to Theorem 9.2.b of \cite{jacod2012probability}, $\mathcal{I}(z_0)$ converges almost surely.
  
  The explicit expression of $\mathbb{E}[G_{\site}(z_0)]$ is not given here, but readers can refer to \cite{rachad20193d} in which the derivation steps are provided. Also, in the case of 2D beamforming with a horizontal radiation pattern and fixed downtilt, the expression of $\mathbb{E}[G_{\site}(z_0)]$ becomes very simple to derive and it is given according to \cite{rachad20193d} by
  
  \begin{equation}
  \mathbb{E}[G_{\site}(z_0)]=\frac{3\eta V(\phi_{\site,c}) \Gamma(\frac{1}{2}-w_h)}{\sqrt{\pi}\Gamma(1-w_h)}.
  \end{equation} 
  
  Now to calculate the average UL to DL $ISR$, we average first over all the cells operating in UL. Then we average over the shadowing log-normal RVs and over the mobiles $z_{\site,c}$ random locations conditionally on $z_0$. It follows that
  {
  \begin{align}
  &\mathbb{E}[ D_{\uparrow}(z_0)]= \frac{9P^* \alpha_u e^{\frac{\ln^2(10)\tilde{\sigma}^2}{200}}}{4 \pi \delta P }  \sum_{c=1}^{3} \sum_{\site \in \Lambda^*} \int_{\vartheta_c-\frac{\pi}{3}}^{\vartheta_c+\frac{\pi}{3}}  \bigg[   \nonumber \\
 &\int_{0}^{\frac{2\delta}{3}\cos^{-2\omega_h}(\phi - \vartheta_c)} \frac{r^{2b} x^{2bk} }{\cos^{-2\omega_h}(\phi - \vartheta_c) |\site - r e^{i\theta}- x e^{i\phi}|^{2b} } dx \bigg] d\phi.
  \end{align}
 }

During the UL cycle of the serving cell $\site_0$, the UL transmitted signal by the mobile $z_0$ is interfered by the DL signal of cells transmitting in DL and also by the signal of mobiles transmitting in UL. We define the UL $ISR$ as

 \begin{equation}
\mathcal{I}_{UL}(z_0)= U_{\uparrow}(z_0) + U_{\downarrow}(z_0)
\label{ISRULdef}
\end{equation}
where $U_{\downarrow}(z_0)$ is DL to UL $ISR$ and $U_{\uparrow}(z_0)$ refers to UL to UL $ISR$.\\

The interfering signal coming from BSs in DL, which have fixed positions, is received at the location of $\site_0$. Hence, by averaging conditionally on $z_0$, over all DL transmitting cells, shadowing and beamforming radiation patterns RVs, $U_{\downarrow}(z_0)$ is expressed by

\begin{align}
&\mathbb{E}[ U_{\downarrow}(z_0)]=\frac{P \alpha_d}{P^*} e^{\frac{\ln^2(10)\tilde{\sigma}^2}{200}} \sum_{\site \in \Lambda^*} |\site-\site_0|^{-2b} r^{2b(1-k)} \mathbb{E}[G_{\site}(\site_0)],
\label{DLtoUL}
\end{align}
with $G_{\site}(\site_0)$ is the 3D beamforming antenna radiation patterns coming from the DL sites and received at the location of $\site_0$.
$\sum_{\site \in \Lambda^*} |\site-\site_0|^{-2b} r^{2b(1-k)}$ is a convergent series on $x=\frac{r}{\delta}$ according to \cite{rachad2018interference} and its expression is given by

\begin{equation}
\sum_{\site \in \Lambda^*} |\site-\site_0|^{-2b} r^{2b(1-k)} = \frac{\omega(b)}{\delta^{2bk}} x^{2b(1-k)}.
\end{equation}
Hence by using the same reasoning as we did for the DL to DL $ISR$, we can show that $U_{\downarrow}(z_0)$ converges almost surely.\\ 

UL to UL interference is generated by mobiles $z_{\site,c}=\site + r_{\site,c}e^{i\theta_{\site,c}}$ in neighboring cells transmitting in UL and also from the UL transmission in the two co-sectors of $\site_0$. Recalling the fact that $r_{\site,c}$ is a RV uniformly distributed in $[0,\frac{2\delta}{3}U(\theta_{\site,c}-\vartheta_c)]$, $\theta_{\site,c}$ is uniform RV in $[\vartheta_c-\frac{\pi}{3}, \vartheta_c+\frac{\pi}{3}]$ and considering the fractional power control model applied to the UL transmission, it follows that 
{
	\begin{align}
	\mathbb{E}&[ U_{\uparrow}(z_0)]=-1 + \frac{9}{4\pi \delta}e^{\frac{\ln^2(10)\tilde{\sigma}^2}{200}} \sum_{\site \in \Lambda} \int_{\vartheta_c-\frac{\pi}{3}}^{\vartheta_c+\frac{\pi}{3}}  \nonumber \\ 
	&\bigg[ \int_{0}^{\frac{2\delta}{3}\cos^{-2\omega_h}(\phi- \vartheta_c)} \frac{| \site + x e^{i\phi}|^{-2b}x^{2bk}}{\cos^{-2\omega_h}(\phi- \vartheta_c) r^{2b(k-1)} } dx  \bigg] d\phi.
	\label{ULtoUL}
	\end{align}}

Based on the expressions of the DL and UL $ISR$ derived previously, we define the DL and UL $SINR$, denoted respectively by $\Pi_{DL}$ and $\Pi_{UL}$, as follows

\begin{align}
&\Pi_{DL}(z_0) = \frac{1}{\eta~\mathcal{I}_{DL}(z_0) + y_{0} x^{2b}} \\
&\Pi_{UL}(z_0) = \frac{1}{\eta~\mathcal{I}_{UL}(z_0) + y^{'}_{0} x^{2b(1-k)}}
\label{SINRDL}
\end{align}
where $y_{0}=\frac{P_{N} a \delta^{2b}}{P \chi(\site_0,z_0) }$, $y^{'}_{0}=\frac{P_{N} a \delta^{2b(1-k)}}{P^{*} \chi(z_0,\site_0) }$, $P_{N}$ is the thermal noise power and $\eta$ is the average load over the interfering cells.\\
 
Finally, we define the coverage probability (CCDF of $SINR$) as the probability that a mobile user is able to achieve a threshold $SINR$, denoted by $\gamma$, in UL and DL transmissions.

\begin{equation}
\Theta(\gamma) = P( SINR > \gamma)
\label{coverage}
\end{equation}

For any scenario of user location distributions, the coverage probability is given by 

\begin{equation}
\Theta(\gamma) = \int_{\site_{0}} \mathds{1}( SINR > \gamma)dt(z_0)
\label{distributioncov}
\end{equation}
such that $\int_{\site_{0}}dt(z_0) = 1$.

\section{Simulation results}
  
  In this section, we simulate in MATLAB the 3D beamforming model considering static and dynamic TDD scenarios. Table. \ref{tab:table1} shows the different parameters used to perform this simulations. 
  
  \begin{table}[h!]
  	\begin{center}

  		\begin{tabular}{l|c|r} % <-- Alignments: 1st column left, 2nd middle and 3rd right, with vertical lines in between
  			
  			%	\textbf{Parameters} & \textbf{Values} \\

  			\hline
  			Macro-cells power $P$ & 43dBm \\

  			\hline 
  			Target power cell specific  $P^*$ & 20dBm \\ 
  			
  			\hline 
  			Noise power  $P_N$ & -93dBm \\ 
  			
  			\hline
  			Number of rings (Macro-cells)  & 5 (90 interfering sites) \\
  			
  			\hline
  			Inter-site distance $\delta$ & 0.75km   \\
  			
  			\hline 
  			Antennas gain & 17.5dBi \\

  			\hline 
  			BSs height $l_b$ & $0.02km$ \\ 
  			
  			\hline 
  			Antenna downtilt & $8\degree$ \\ 
  			
  			\hline 
  			Shadowing standard deviation & 6dB \\ 
  			
  			\hline
  			Propagation factor $a$ & Outdoor: 130dB    \\
  			
  			\hline
  			System bandwidth & Macro-cells:20Mhz  \\
  			\hline
  			
  			Path loss exponent $2b$ & 3.5  \\
  			\hline
  			
  		\end{tabular}
  		\caption{Simulation parameters.}
  		\label{tab:table1}
  	\end{center}
  \end{table}
  
\begin{figure}[tb]
  	\centering
  	\includegraphics[height=5.5cm,width=9cm]{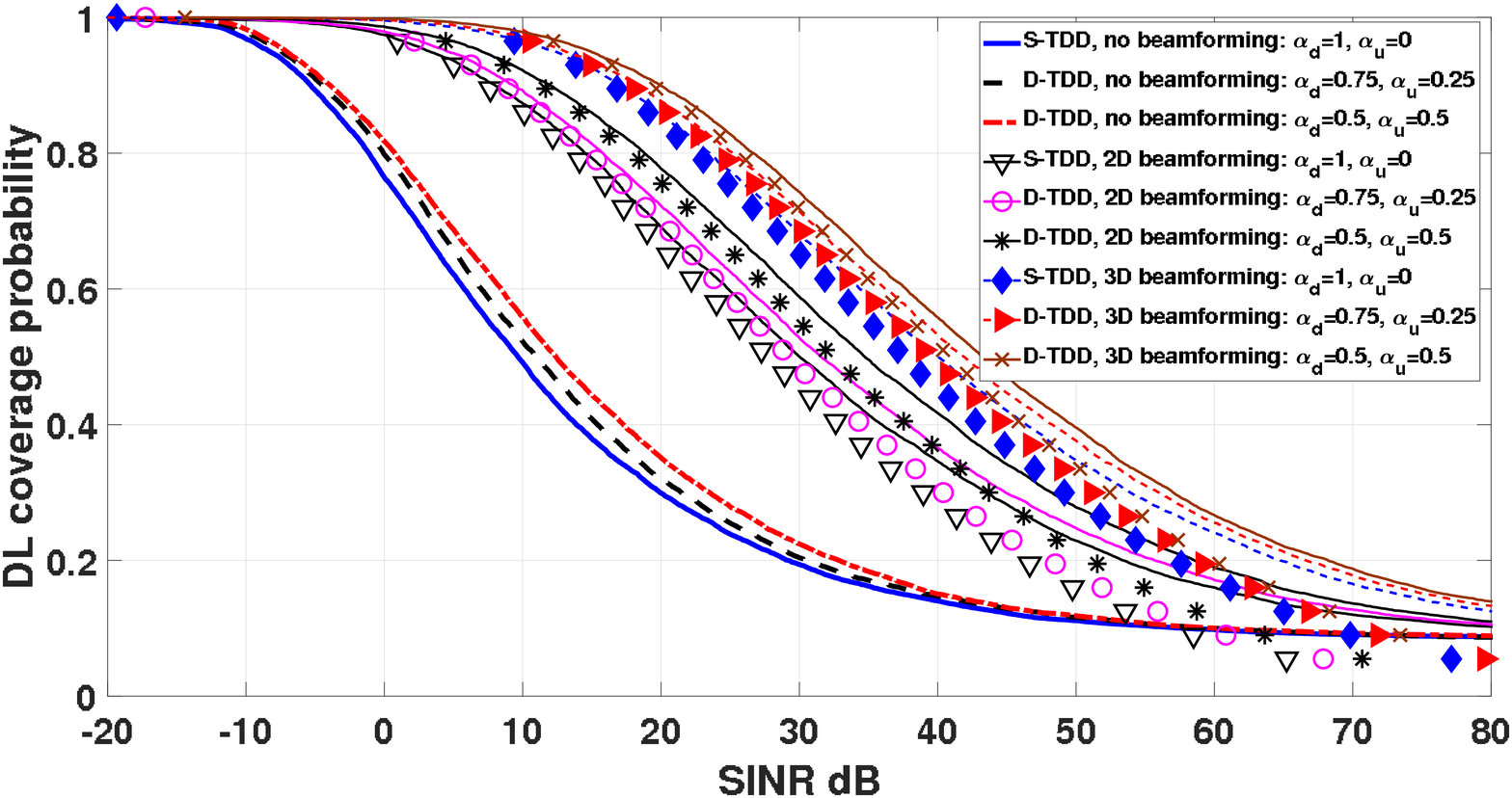}
  	\caption{ DL Coverage probability ($2b=3.5$, $k=0.4$).}
  	\label{cp DL1}
  \end{figure}

  We plot in Fig. \ref{cp DL1} the coverage probability curves in DL obtained by using Monte Carlo simulations for 20000 mobile locations $z_0$ uniformly distributed. We compare the static TDD and Dynamic TDD considering three scenarios: without beamforming scheme, with 2D beamforming and 3D beamforming. Starting from a static TDD configuration where all the sites $\site$ are transmitting in DL, Fig. \ref{cp DL1} shows that the coverage probability increases when D-TDD is activated,  with $\alpha_{d}=75\%$ and $\alpha_{d}=50\%$, for the three scenarios. This behavior is expected since the macro-cells BSs transmit with high power level and generate strong interference compared to interfering mobiles $z$ transmitting in UL. In the second scenario, we contemplate 2D beamforming where only the horizontal radiation pattern with a half power beam-width $\theta_{h3dB}=14\degree$ and a fixed antenna downtilt are considered. As expected, there is an important enhancement in system performance, translated by an increase in the coverage probability, for both S-TDD and D-TDD, compared to the first case. This enhancement becomes more obvious in the third scenario when 3D beamforming is implemented with $\theta_{h3dB}=14\degree$ and a vertical half power beam-width $\theta_{v3dB}=8\degree$.  Actually, most 4G  BSs use a linearly arranged array of antennas placed at the top of BSs. Recognizing the difficulty to increase the number of antennas because of size constraints, FD-MIMO based on a 2D array of antennas offers the possibility to increase their number. Also, it provides the capability to adapt dynamically beam patterns in the horizontal and vertical dimensions. Given that D-TDD is more convenient with DL transmission direction, adding 3D beamforming further improves performance without losing the flexibility in resource assignment.\\ 
  
  \begin{figure}[tb]
  	\centering
  	\includegraphics[height=5.5cm,width=9cm]{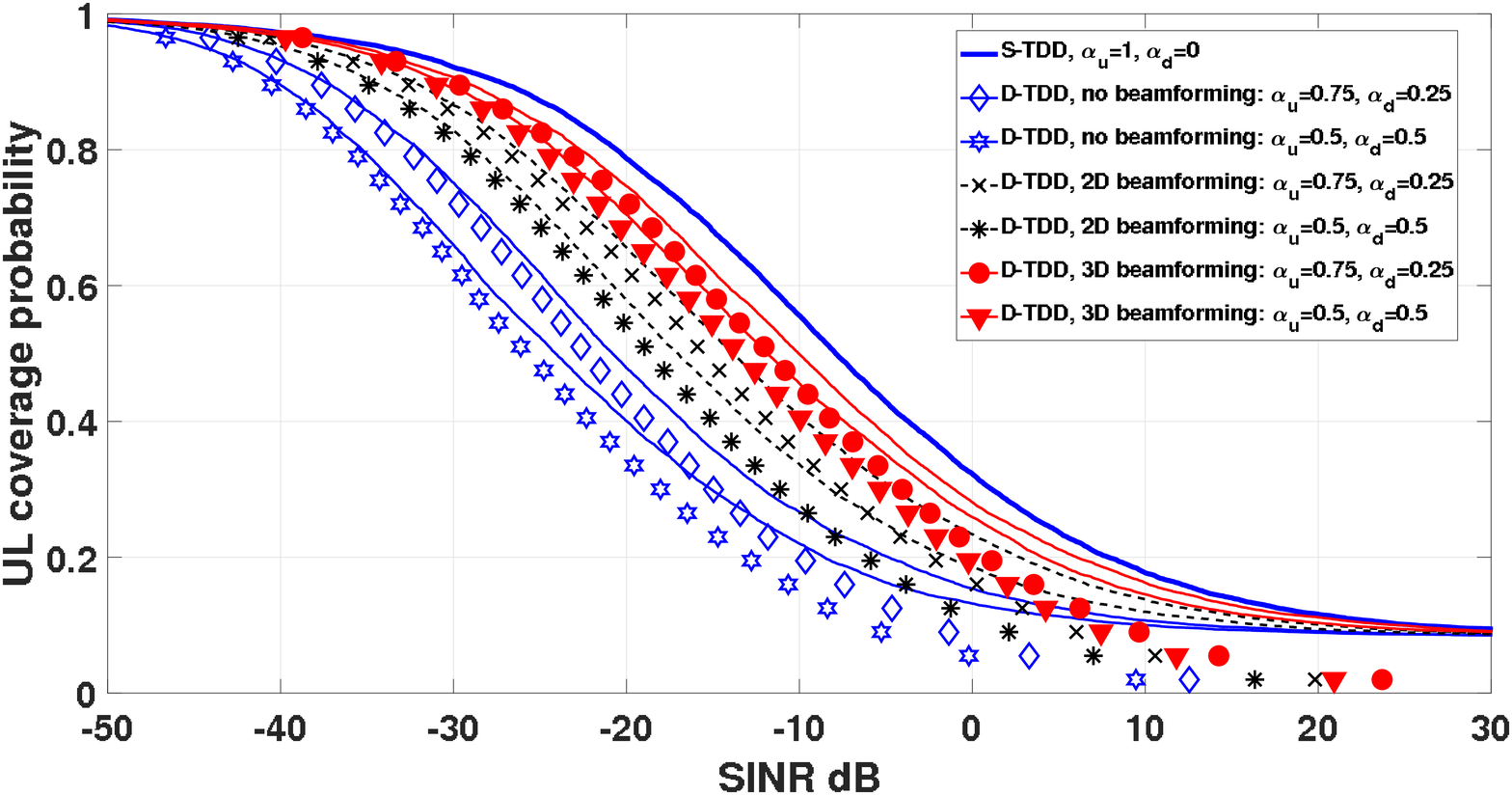}
  	\caption{ UL coverage probability ($2b=3.5$, $k=0.4$).}
  	\label{cp UL1}
  \end{figure}
  
  To analyze the system behavior during the UL transmitting cycle of $\site_0$, we plot in Fig. \ref{cp UL1} the UL coverage probability considering the same scenarios as in DL. It is worth mentioning that for S-TDD UL transmission, there is no beamforming mechanisms since all BSs are in UL and mobiles are equipped with omni-directional antennas. Hence, we consider only one S-TDD scenario and we compare it to the three scenarios raised in the previous paragraph. The main obvious observation from Fig. \ref{cp UL1} is that without beamforming schemes, the coverage probability undergoes a huge degradation when the system switch from the static configuration to the dynamic one, with $\alpha_u=75\%$ and $\alpha_u=50\%$. This degradation is mainly coming from the strong interfering signals of DL BSs that make the system very limited if no interference mitigation schemes are set up. It is noteworthy to mention that the results shown in Fig. \ref{cp DL1} and Fig. \ref{cp UL1} are in agreement with theoretical results provided in \cite{rachad2018interference} and simulation results of \cite{khoryaev2012performance}. Moreover, with 2D beamforming, one can observes from Fig. \ref{cp UL1}  that there is an enhancement in the UL coverage probability when D-TDD is activated compared to the static configuration. This enhancement becomes more significant when 3D beamforming is considered for DL BSs and in this case, the UL coverage probability approaches the one of S-TDD.\\

\begin{figure}[tb]
	\centering
	\includegraphics[height=5.5cm,width=9cm]{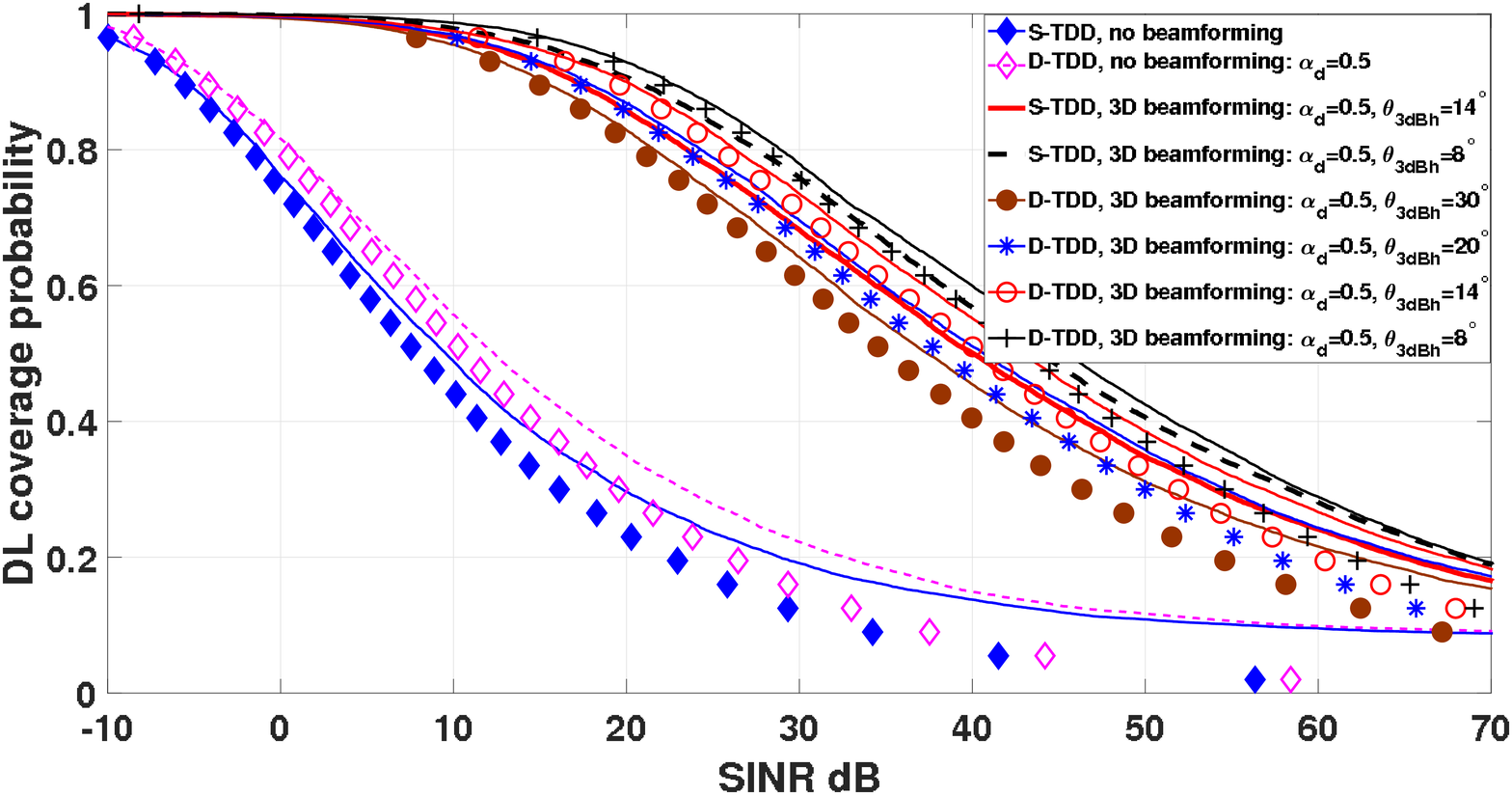}
	\caption{ DL coverage probability: D-TDD vs S-TDD with 3D beamforming considering different half power beam-widths ($2b=3.5$, $k=0.4$).}
	\label{ouverture DL}
\end{figure}

\begin{figure}[tb]
	\centering
	\includegraphics[height=5.5cm,width=9cm]{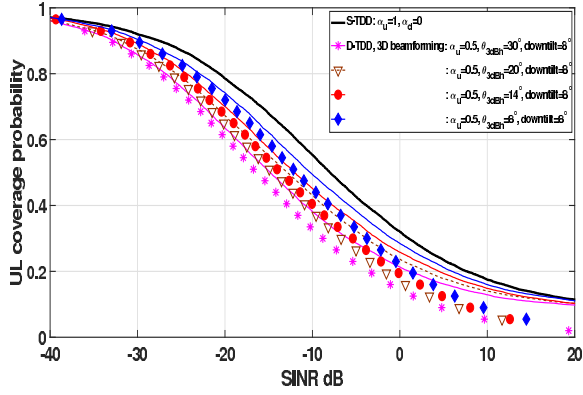}
	\caption{UL coverage probability: D-TDD vs S-TDD with 3D beamforming considering different half power beam-widths ($2b=3.5$, $k=0.4$).}
	\label{ouverture UL}
\end{figure}

In Fig. \ref{ouverture DL} and Fig. \ref{ouverture UL} we plot respectively the DL and UL coverage probability curves of a D-TDD scenario combined with 3D beamforming. We consider different horizontal half power beam-widths ($\theta_{h3dB}=30\degree, 20\degree, 14\degree, 8\degree$) and a vertical half power beam-width $\theta_{v3dB}=8\degree$. When the serving site $\site_0$ is operating in DL,  it can be observed that the DL coverage probability increases as the beam width decreases for both D-TDD and S-TDD. Also, one can notice that the gain obtained with D-TDD is quite important than S-TDD without beamforming. Moreover, the coverage probability of a S-TDD network with 3D beamforming having small $\theta_{h3dB}$ approaches the one of a D-TDD system. Actually, the beam-width is related to the number of transmit antennas used by BSs. When this number increases, the signal is focused on a specific zone of the cell. Hence, interference coming from neighboring sites are reduced significantly. This leads to an enhancement of $SINR$ and thus an enhancement of the coverage probability. Similar results are observed for the UL scenario as shown in Fig. \ref{ouverture UL}. With 3D beamforming based D-TDD, as the horizontal half power beam-width is reduced, the UL coverage probability is enhanced and approaches the one of S-TDD. This means that 3D beaforming reduces significantly the DL to UL interference and make D-TDD feasible for macro-cells' network.\\

\begin{figure}[tb]
	\centering
	\includegraphics[height=5.5cm,width=9cm]{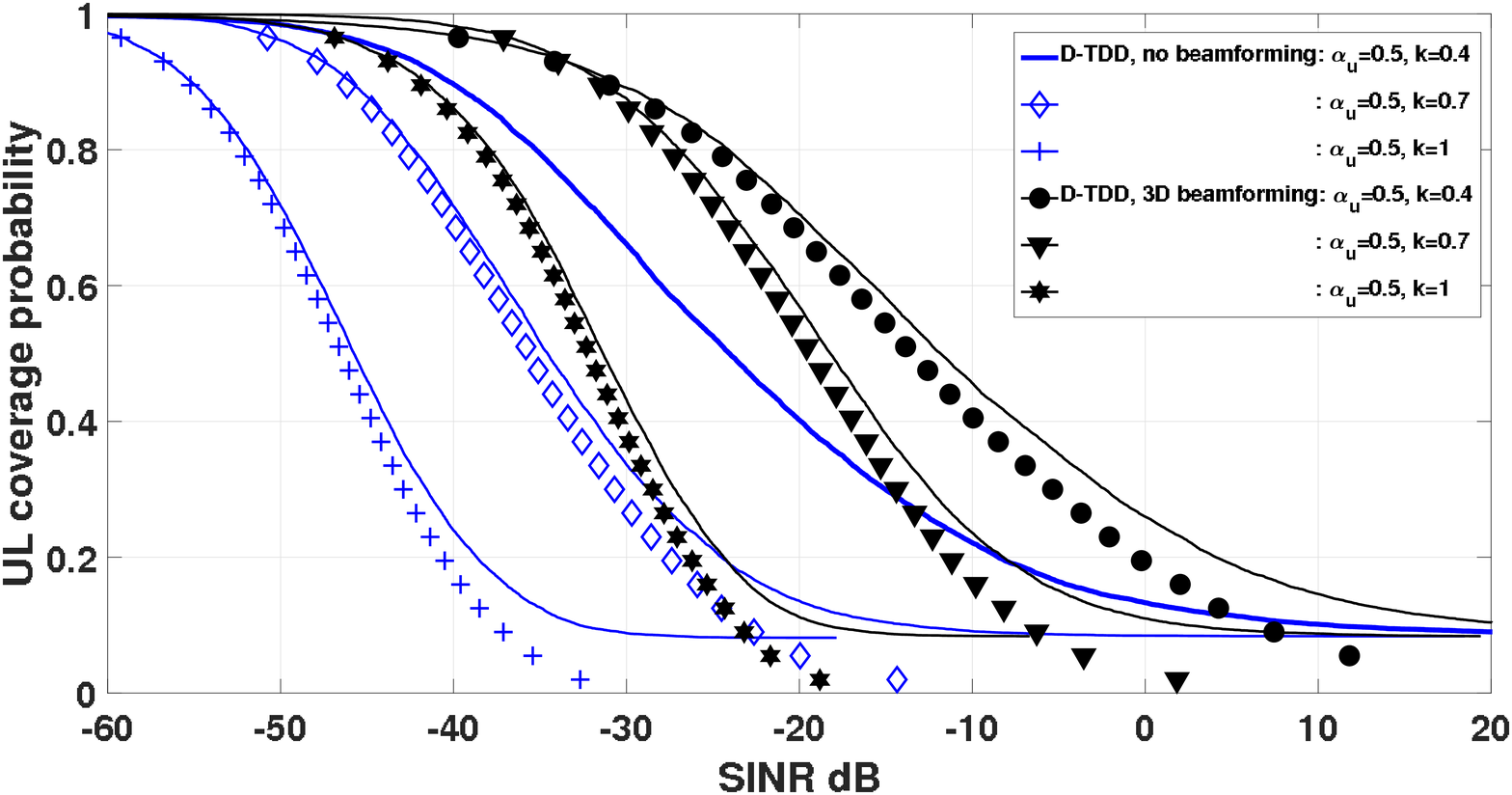}
	\caption{ UL coverage probability: fractional power control effect: $2b=3.5$, $k=0.4$.}
	\label{FCP UL}
\end{figure}

Finally, to analyze the effect of the fractional power control considered for the UL transmission direction, we plot in Fig. \ref{FCP UL} the UL coverage probability of a D-TDD system, with and without 3D beamforming, considering different FPC factor values ( $k=0.4$, $k=0.7$ and $k=1$). One can notice that the coverage probability is decreasing as the FPC factor is increasing, for the both scenarios. Actually, FPC aims at providing the required $SINR$ to UL users while controlling at the same time their interference. When FPC factor $k=1$ the path loss is completely compensated and the cell-specific target power $P^*$ is reached. Thus, the interference coming from mobiles $z$ in UL is higher especially if a mobile is located in the edge of a neighboring cell. When FPC factor $0<k<1$, the scheme indemnifies partially the path loss. The higher is the path loss the lower is the received signal. This means that there is a compromise between the path loss and the $SINR$ requirements. Therefore, interference are likely to be controlled, which explain the enhancement of the coverage probability.\\

\section{Conclusion} 
  
  In this paper, we have proposed a 3D beamforming based D-TDD interference mitigation scheme where antenna horizontal and vertical radiation patterns depend on the spatial distribution of users' locations. We have explicitly analyzed the ISR metrics covering different interference scenarios. Through system level simulations, we have shown that D-TDD is in favor of DL transmission and the gain is important when 3D beamforming is considered. For the UL transmission, we have shown that the system is very limited by the strong interference coming from DL transmitting BSs. With 3D beamforming this interference is significantly reduced and D-TDD UL transmission performance approaches S-TDD. Also, we have shown that DL to UL ISR and DL to DL ISR are two almost sure convergent series of independent RVs. The almost sure  convergence is an important result since it implies convergence in probability and thus convergence in distribution. Further extension of this work could include the analysis of beam coordination mechanisms in UL transmission direction to reduce UL to DL and UL to UL interference since this interference is difficult to deal with because mobiles move around randomly.\\

%   Conditionally on the mobile location $z_0$, by averaging $\mathcal{I}_{DL}(z_0)$ we obtain
%   
%   \begin{align}
%   \mathcal{AI}_{DL}(z_0)=-1 + \alpha_d e^{\frac{\ln(10)\tilde{\sigma}^2}{5}} \mathbb{E}[G_{\site}(z_0)] \sum_{\site \in \Lambda} \mathbb{E}[G_{\site}(z_0)] |\site - z_0|^{-2b} r^{2b} +
%   \end{align}  

\balance
\bibliographystyle{IEEEtran}
\bibliography{IEEEabrv,TelecomReferences}

\end{document}